# Emerging trends on the topic of Information Technology in the field of Educational Sciences: a bibliometric exploration

Carlos Luis González-Valiente
Grupo Empresarial de la Industria Sidero M…

# EMERGING TRENDS ON THE TOPIC OF INFORMATION TECHNOLOGY IN THE FIELD OF EDUCATIONAL SCIENCES: A BIBLIOMETRIC EXPLORATION

Carlos Luis González-Valiente
Grupo Empresarial de la Industria Sidero Mecánica,
Department of Informatics and Management of Information
[carlos.valiente@fcom.uh.cu]

## ABSTRACT

The paper presents a bibliometric analysis on the topic of Information Technology (IT) in the field of Educational Sciences, aimed at envisioning the research emerging trends. The ERIC data base is used as a consultation source; the results were subjected to productivity by authors, journals, and term co-occurrence analysis indicators for the period 2009-2013. The productivity of Computers & Education, and Turkish Online Journal of Educational Technology-TOJET, as well as the preceding authors from Canada, have been emphasized. The more used terms are the following: Information technology, foreign countries, educational technology, technology integration, and student attitudes. Researches performed here seem to have a largely qualitative character, highlighting computers and internet as the mostly explored technological objects. The largest subject matter trend refers to the integration of IT in the higher education learning context, and its incidence over the teaching methods.

**Keywords:** Information technology; Educational Sciences; Bibliometrics; Scientific production; Research trends; ERIC database

## 1. - INTRODUCTION

The technological revolution emerged in the late 20th century has brought about a re-dimensioning process of the theoretical and practical ways of thinking in the disciplines fields. In the case of Educational Sciences (ES), the information technologies (ITs) have open new possibilities to teaching (Yusuf, 2005; Gómez, 2012), which has implied a re-formulation of teaching-learning process's practical methods (Reddy, 2006). The application of technologies has undoubtedly become a paradigmatic factor for all fields of knowledge. Formosinho, Reis & Renato (2013, p. 50), when analyzing specifically ES from an operational and technological dimension, state that: "the technocentric thinking brings with it a composite reductionist vision of human life that encompasses a model of society, an idea of education, and even a conception of knowledge, where the essential value lies in a narrowed understanding of «usefulness»".

ITs are particularly conceptualized as tools that effectively support teaching, learning, and education innovation management, thus contributing towards the improvement of educational efficiency and quality (Peeraer & van Petegem, 2012). Such peculiarities make them to be considered also as: (a) a tool for addressing challenges in teaching and learning, (b) a change agent, and (c) a central force in economic competitiveness (Yusuf, 2005). However, studies on technologies integration in educational realm have been conditioned by multiple elements (Costa, 2007). Empiric results reveal evidence as to the benefits of the ITs application in generating knowledge in students (Mcanally-Salas, Navarro-Hernández & Rodríguez-Lares, 2006; Sáez & Ruiz, 2012); a fact that has contributed to the prevailing need to train professors (Rangel & Peñalosa, 2013; Sieiro, 1994; Boza, Tirado & Guzmán-Franco, 2010).

Thereof, as part of the innovation activating, many educational centers have focused themselves in monitoring useful and timely information technologies, as to teaching. Even though, beyond the technological monitoring, there is a phenomenon which, marked by research, discloses a high scientific production on the ITs topic within the ES realm.

### *1.1. - Literature review*

Regarding the studies oriented towards bibliometric characterization of the scientific activity associated to ITs in ES, the work by Costa (2007) is to be highlighted; he investigated educational technologies research behavior, as to the master's thesis discussed in Portugal, in the period between 1960 and 2005. Here, Costa identified ITs as the main topic dealt with; a pattern that has also been visible in the findings by Ozarslan & Balaban-Sali (2012), Rodríguez & Remón (2014) and Potvin & Hasni (2014). Such results are, in a parallel manner, the detonators in the studies by Assefa & Rorissa (2013), who, through a terms co-concurrence analysis, characterized the STEM education realm to identify the ideas that have implications on the curricular development. Johnson & Daugherty (2008) devote themselves to assessing the quality and the characteristics of educational technology research in the period 1997-2007.

On the other hand, Martin *et al.* (2011) likewise analyzed, through bibliometric techniques, the technologies that have suited the educational systems, as well as their degree of evolution and maturity. With a peculiarity much more reduced to the Vietnamese context, Peeraer & van Petegem (2012) explored the phenomenon of the integration of ITs to teaching, fostering a descriptive analysis towards measuring the incidence level of such technologies in the formation activity. Liu, Wu & Chen (2013), have recently examined the Learning Technologies' (a.k.a. IT) trends in special education, as from 26 studies published in indexed journals (2008-2012). Such enquiring had a dual direction, one directed towards detecting the methodological aspects related with the way of studying such subject matter, and the other, towards focusing on perceiving the typology of the used technology in this field of special education. Research on IT in the educational realm has been taking a quite advanced position regarding other disciplines, not only within ES, but also in the generic context of Social Sciences (Cabero, 2004).

Up to now, there is no ample evidence of the particular exploration of scientific production on the research associated to ITs in ES. That is why this article aims at examining, based on the bibliometric methodological principles, the related scientific productivity. To such end, some metric indicators will be used to facilitate the explanation and visualization of research trends in ERIC database within the last publication period (2009-2013), for the latter specializes itself in educational subject matters.

## 2. - MATERIALS AND METHODS

The analysis of the investigation's results in scientific production as to bibliometric indicators, has acquired an especial relevance, for they provide timely characterizations of the different scientific profiles (Miguel, Moya-Anegón & Herrero-Solana, 2006). ES, in particular, is a field that has not been ignorant of this type of perspective analysis, something that can be proven in articles by Phelan, Anderson & Bourke (2000); Dees (2008); van Aalst (2010); and Diem & Wolter (2013). It is also evident that the respective bibliometric indicators contribute to examine the knowledge development and flow, based on the research that has been mostly extracted from ample coverage international databases (Katz, 1999). Besides, it is precisely Bibliometrics as a discipline the one that contributes to the organization of scientific sectors as from sources that facilitate the identification of trends (Spinak, 2001).

### *2.1. - Data source*

ERIC (Education Resources Information Center) database is sponsored by the Institute of Educational Sciences of the US Department of Education, and it is a digital library that indexes over 600 scientific journals on education and information research. It includes bibliography of articles and journals from other sources (books, research synthesis, conference papers, technical reports, policy papers, and other education-related materials), dating back to 1996 to the present (http://ies.ed.gov/ncee/projects/eric.asp).

ERIC facilitates the search and filtering of documents regarding elements such as publication date, descriptor, source, author, publication type, education level, and audience. Besides, it helps to determine that the initial search can be referred to studies whose character can be evaluated, or not, by peers (http://eric.ed.gov/?advanced). It is necessary to bring out that this database has become the object of important bibliometric studies for the educational field (e.g.: Edyburn, 2001; del Mar & Pérez, 2008; Strayer, 2008; Assefa & Rorissa, 2013; Potvin & Hasni, 2014).

### *2.2. - Data gathering and processing*

Information technology was a term defined to search strategy within the title of journal' articles in the period 2009-2013. Such temporary coverage was considered timely, for the goal is to show the most emerging trends on the topic. In order to determine and visualize research trends, several bibliometric indicators fitting publication analysis were applied (Spinak, 2001; Schneider, 2006), such as:

- *Productivity by author*: determined as from distribution of authors by article, disregarding its role as main or secondary author.

- *Productivity by journal*: determined as from distribution of articles by journals.
- *Term co-occurrence analysis*: determined as from key words declared in articles. Those descriptors co-occurring twice or more were the only ones used.

It is necessary to clear out that ERIC provides quantitative results in searches as from statistic counting; but in the descriptors' case, there are no relations among them offered, and this is an important element for the analysis of the terminological co-occurrence. That is why it was necessary to use EndNote software (X4 version, www.endnote.com), designed by Thompson Reuters company for the management and normalization of bibliographic registries. They were exported in a *.txt* format file, to be later on used by the Bibexcel tool, developed by Olle Persson (www8.umu.se/inforsk/Bibexcel), that has the applications for the bibliographic data analysis that would later on served to generate maps which illustrated networks of relations among the terms. With the use of Bibexcel, a *.net* file was created compatible for visualizing such maps as from the VOSviewer 1.4.0 program; whose specialization is based on the creation, visualization, and exploration of science bibliometric maps (www.vosviewer.com).

## 3. - RESULTS AND DISCUSSION

After the search, 142 results were obtained, out of which 61% of the articles are Report-research type, while 21% are Report-evaluation, and 17% Reports-descriptive. Regarding the educational level, 47% corresponds to higher education, which is similar to Hwang & Tsai's (2011) results, a study whose methodological platform was likewise based on metrics. There are other levels: Postsecondary (16%), Elementary secondary education (12%), High schools (12%), and Adult education (8%).

Author's productivity does not surpass three articles per author, and most of them are originated in the university environment (99%), a common pattern with Costa's findings (2007). The strong presence of such institution evidences the criterion which states that the scientific research generally emerges from higher education sector. Canadian authors have a strong presence; they sum up 40% of all 15 articles presented in Table 1. It is precisely Canada the country that, according to SCImago Journal & Country Rank[i] in the educational field, is ranked fourth among countries as to the level of productivity (1996-2012), with a total of 8,302 documents that equals 71 H index (http://www.scimagojr.com/countryrank.php?area=3300&category=3304&region=all&year=all&order=it&min=0&min_type=it).

**Table 1**. Prolific authors.

| Author | Country | Institution | # of articles |
|---|---|---|---|
| Rhonda Amsel | Canada | McGill University, Department of Psycology | 3 |
| Jennison V. Asuncion | Canada | McGill University, Adaptech Research Network | 3 |
| Jillian Budd | Canada | Dawson College, Adaptech Research Network | 3 |
| Catherine S. Fichten | Canada | McGill University, Department of Psychiatry | 3 |
| Jef Peeraer | Vietnam | Flemish Association for Development Cooperation and Technical Assistance | 3 |
| Peter Van Petegem | Belgium | University of Antwerp, Institute for Education and Information Sciences | 3 |
| Maria Barile | Canada | Dawson College, Adaptech Research Network | 2 |
| Betty Breed | South Africa | North-West University, School of Natural Science and Technology for Education | 2 |
| Mercedes Gonzalez-Sanmamed | Spain | Universidad de A Coruña, Facultade de Ciencias de la Educación | 2 |
| Tony Koppi | United States | Goshen College, Informatics | 2 |
| Elsa Mentz | South Africa | North-West University, School | 2 |
| | | of Natural Science and Technology for Education | |
| Mai Nhu Nguyen | Canada | Dawson College, Adaptech Research Network | 2 |
| Hatice Ferhan Odabasi | Turkey | Anadolu University, Education Faculty, Computer and Instructional Technologies Education Department | 2 |
| Albert Sangra | Spain | Universitat Oberta de Catalunya, eLearn Center | 2 |
| Grace Tan | Australia | Victoria University Melbourne, College of Engineering and Science | 2 |

The most productive journals have been Computers & Education (United Kingdom), and Turkish Online Journal of Educational Technology-TOJET (Turkey). Each of them contain, proportionally distributed, 11 articles. Computers & Education in particular has been validated, within the sample defined by Keser, Usunboylu & Ozdamli (2011), as the most published journal (from 2005 to 2010) on technologies supporting collaborative learning. US journals have also been prolific (see Table 2). Regarding authors and journals, the North American region seems, likewise, very productive, a statement confirmed by Barth & Rieckmann (2013), Potvin & Hasni (2014) and SCImago Journal & Country Rank (2014) itself.

**Table 2**. Most productive journals.

| **Journal** | **Country** | **# of articles** |
|---|---|---|
| Computers & Education | United Kingdom | 11 |
| Turkish Online Journal of Educational Technology-TOJET | Turkey | 11 |
| Journal of Information Technology Education | United States | 8 |
| EDUCAUSE Review | United States | 4 |
| Educational Sciences: Theory and Practice | Turkey | 4 |
| Journal of Information Systems Education | United States | 4 |

The term co-occurrence analysis, allows to unveil the subject matter's interconnections which are more intense, or not, regarding frequency of main keywords in studies. According to Table 3, those most co-occurring are Information technology, Foreign Countries, Educational technology, Technology integration, Student attitudes, etc.

**Table 3**. List of the most co-occurring keywords.

| **Term** | **Co-occurrence** |
|---|---|
| Information Technology | 114 |
| Foreign Countries | 90 |
| Educational Technology | 45 |
| Technology Integration | 29 |
| Student Attitudes | 25 |
| Internet | 24 |
| Computer Uses in Education | 24 |
| Teaching Methods | 23 |
| Questionnaires | 22 |
| Technology Uses in Education | 22 |
| Higher Education | 21 |
| Interviews | 21 |
| Teacher Attitudes | 20 |

Figure 1 presents a map showing the keywords' relationship, which have been grouped in 10 main clusters. The difference among each of them is shown through color shades; the size of items is presented through the level of frequency, while those in the peripheral space show the degree of approach to the main topic (Information technology).

**Figure 1**. Co-occurrence map of ERIC's keywords

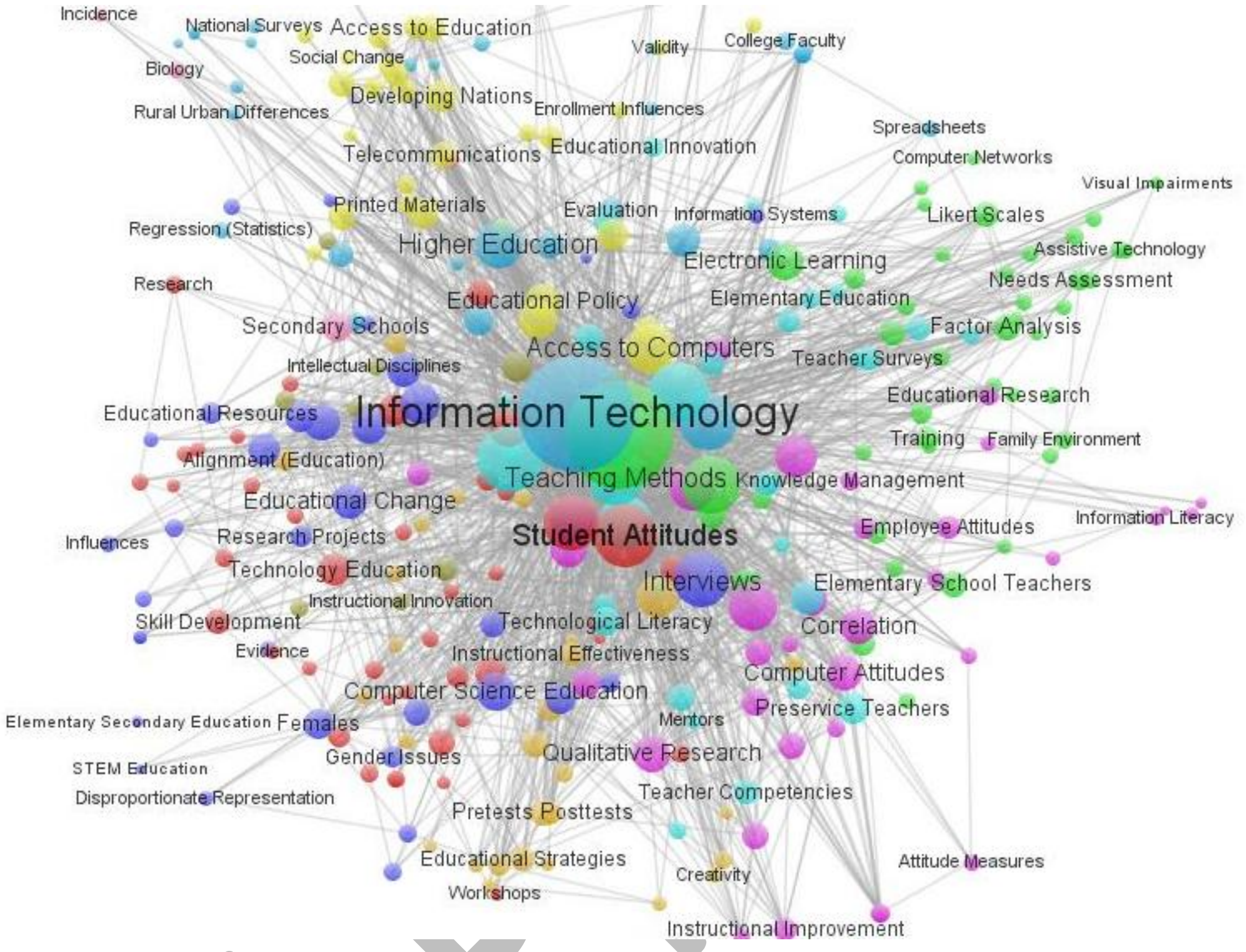

In order to timely analyze the most distinctive clusters, a second map (see Figure 2) was developed, showing a thick view of every cluster, as well as their interconnection's levels. Cluster 1 (in red) is made up by 45 items, being *Students attitudes* the most intense, and which has a higher links strength (ls) with *Teacher attitudes* (ls: 6), *Teaching methods* (ls: 6), and *Interviews* (ls: 4) keywords. Other similarly relevant, though less intense terms, are *Technology uses in education*, *College students*, *Technology education*, *Undergraduate students*, *College instruction*, *Learning processes*, *Instructional design*, and *Skill development*. This cluster concentrates queries carried out as from learning perspective; all of which have been demonstrated after the presence and association of keywords surrounding *Active learning*, *Adult learning*, *Blended learning*, *Computer assisted instruction*, *Constructivism (learning)*, *Cooperative learning*, and *Experiential learning* subject matters.

Cluster 2 (in green) is made up by 45 items, being *Foreign countries* the most co-occurring one. Its larger link strength is given through *Information technology* (ls: 68), *Educational technology* (ls: 34), *Computers uses in education* (ls: 20) and *Teaching methods* (ls: 17). There are other categories similarly intense, such as *Questionnaires*, *Electronic learning*, *Observation*, *Elementary school teachers*, *Classroom techniques*, *Factor analysis*,

*Linkert scales*, and *Needs assessment*. This cluster's less co-occurring terms correspond to methodological matters.

**Figure 2**. Cluster density view of keywords co-occurrence

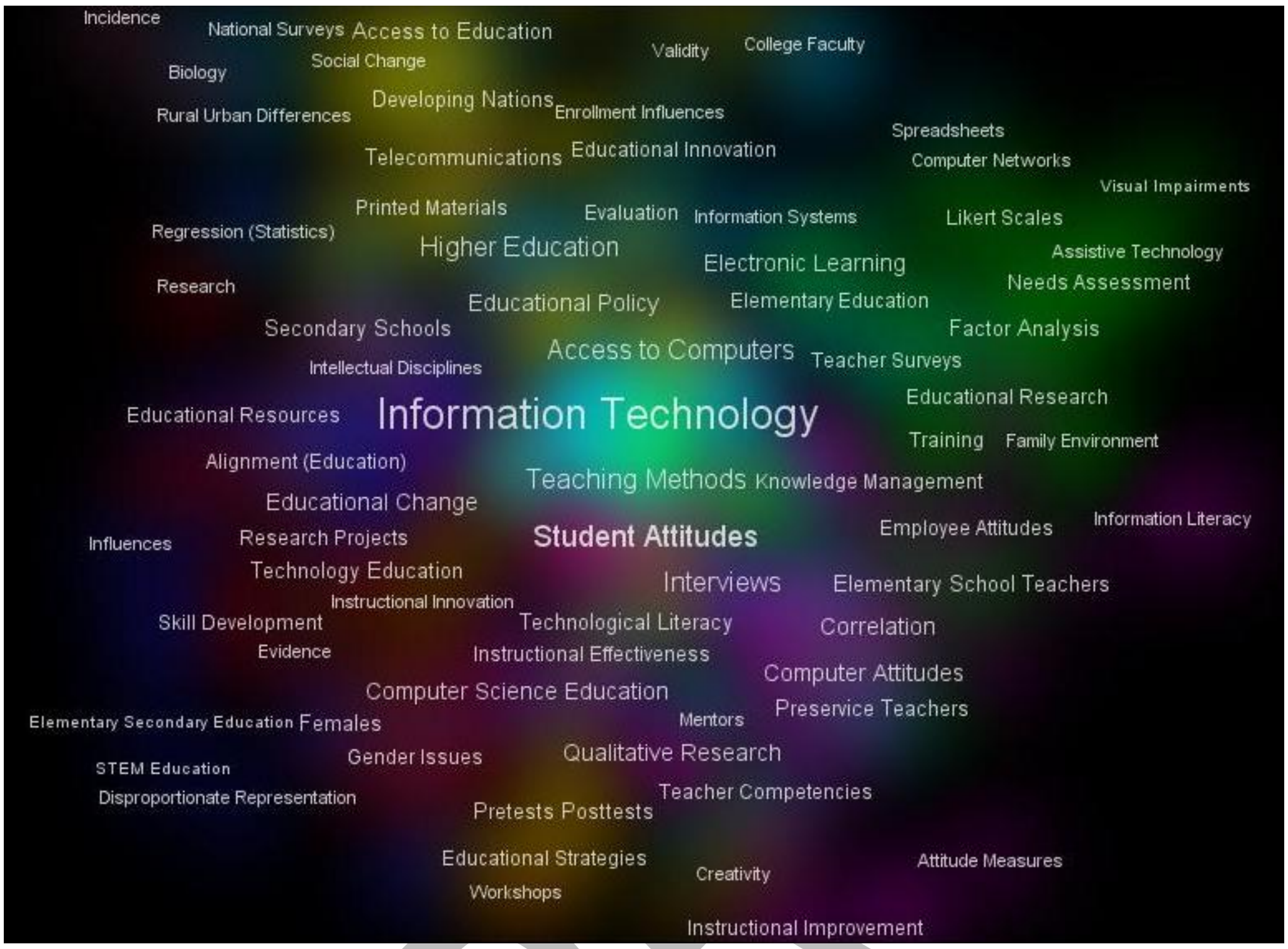

In cluster 3 (in purple, 38 items) *Interviews* is the more intense term, maintaining solid relations with *Student attitudes* (ls: 4), *Computer literacy* (ls: 3) and *Observation* (ls: 3). Other intense terms are *Program effectiveness*, *Performance factors*, *Comparative analysis*, *Educational change*, *Computer science education*, and *Comparative education*. Such cluster reveals the ideas associated with exploration of transformation elements and evolution in educational field. On the other hand, cluster 4 (in yellow), made up by 34 terms, pretends to refer to social and normative matters of educational interest. The most co-occurring term has been *Access to computers*, closely linked with *Foreign countries* (ls: 14), *Educational technology* (ls: 7), *Computers uses in education* (ls: 4) and Internet (ls: 4). Here, *Influence of technology*, *Educational technology*, *Distance education*, *Program implementation*, *Developing nations*, *International education*, *Educational development*, and *Access to education* are also highlighted.

Cluster 5 (in dark pink, 34 items), are made up by terms associated to computing practices. *Computer Literacy* is the most co-occurring keyword, strongly linked with *Self efficacy*, *Undergraduate students*, and *Interviews*. In a lower level, *Case studies*, *Computers attitudes*, *Qualitative research*, *Correlation*, *Predictor variables*, and *Computers mediated communication* are highlighted. In addition, cluster 6's 33 items (in navy) are highlighted for standing closer to the center of the map. The highest intensity resides in *Educational technology* category, whose strongest relations are *Foreign countries* (ls: 34), *Information technology* (ls: 29), *Technology uses in education* (ls: 19), and *Teaching* *methods* (ls: 15). Other relevant methods are *Technology integration*, *Computers uses in education*, *Teaching methods*, *Teacher attitudes*, and *Technological literacy*. This cluster reveals the possible subject matters associated with the adoption and integration of IT. Finally, in cluster 7 (sky blue, 30 items) *Information technology* prevails in the center of the map, strongly relating itself *Foreign countries* (ls: 68), *Technology integration* (ls: 20), and *Teaching methods* (ls: 14) subject matters. The meaningful presence of *Internet*, *Higher education*, *Surveys*, *Models*, *Computers*, and *Curriculum development* cannot be precluded.

Analyzing terms from other point of view, as from the methodological perspective, *qualitative research* (7) has been strongly quoted, a pattern that seems to be common in educational field studies (Costa, 2007; Ozarslan & Balaban-Sali, 2012); while as part of methods and techniques, it is necessary to quote *questionnaires* (22) and *interviews* (21) as the most frequent, following next *case studies* (11), *comparative analysis* (9), *use studies* (9), *surveys* (8), and *factor analysis* (5). These methods and techniques have been similarly common in Costa (2007); Barth & Rieckmann (2013); Liu, Wu & Chen (2013); and Potvin & Hasni (2014) findings. In a lower context *content analysis* (3), *regression* (*statistics*, 3), and *multivariate analysis* (2) can be quoted. All of this proofs the multiplicity of methodological positions for the development of meta-analysis which this type of topic requires (Cabrero, 2004).

*Internet* (24) and *computers* (24), objects considered IT means, are highlighted as leaders; while *web sites* (4), *open access technology* (3), *information systems* (3), *video* (3), *multimedia* (2), *electronic libraries* (2), *electronic mail* (2), and *videoconferencing* (2) reached lower positions. The first objects were characterized as being part of the virtual environment already mentioned by Martin *et al* (2011) in his predictive study for the period 2008-2014, when he evaluated the use of the information technologies to be applied in education, and declared by Horizon Report. This same idea is also stated by Keser, Usunboylu & Ozdamli (2011), while examining the application of technologies in collaborative learning.

## 4. - CONCLUSIONS

The present article has provided a not too thorough view of IT in ES, which can be of great interest for the future practical and disciplinary development of such field. Most part of works on the topic are developed within the Anglo-Saxon context; however, the main subject matters referred to here are regarded to have an international scope. This evidences a transversal line in ways of thinking that go beyond specific national contexts. Studies associated with ITs integration in formation stand as a high prerogative, and its relations with teaching methods, in which student-professor-learning context relation is highly implicit.

The level of higher education has been the main context for the study of these subject matters, and a larger approach to other levels of teaching is considered necessary. Though the present analysis was only limited to keywords, and not to all the contents of the articles, the qualitative research, whose methods and techniques play the part of the very descriptive character of the results, have been distinctive. Computers and Internet consolidates themselves as ITs main objects, a highly corresponding element with the present conditionings of information society, and the contents self-management factors.

### 4. - ACKNOWLEDGEMENTS

I would like to thank the editor of journal of Education in the Knowledge Society (EKS) for his consent as to the translation of the present article. I would also like to thank the translation service section of the University of Havana Main Library, and specially to Laura Monteagudo for her excellent work.

### 5.- BIBLIOGRAPHY

---

[i] SCImago Journal & Country Rank is a portal covering journals' and countries' scientific indicators, developed as from the information contained in Scopus® (Elsevier B.V.). Indicators can be used to measure and analyze scientific realms (http://www.scimagojr.com/index.php).